\documentclass[article,useAMS,usenatbib]{mnras}
\usepackage{txfonts}
\usepackage{natbib}
\usepackage{graphicx}
\usepackage{dcolumn}

\begin{document}

\title[Rapid dust formation in the ISM]{Galactic dust evolution with rapid dust formation in the interstellar medium due to hypersonic turbulence}

\author[Mattsson]{Lars Mattsson$^{1}$\thanks{E-mail: lars.mattsson@su.se}}

\pagerange{\pageref{firstpage}--\pageref{lastpage}} \pubyear{2018}

\maketitle

\label{firstpage}

\date{\today}

\begin{abstract} 
Turbulence can significantly accelerate the growth of dust grains by accretion of molecules. For dust dynamically coupled to the gas, the growth rate scales with the square of the Mach number, which means that the growth timescale can easily be reduced by more than an order of magnitude. The limiting timescale is  therefore rather the rate of molecular cloud  formation, which means that dust production in the ISM can rapidly reach the levels needed to explain the dust masses observed at high redshifts. Thus, turbulence may be the solution to the replenishment problem in models of dust evolution in high-redshift galaxies and explain the dust masses seen at $z = 7-8$. A simple analytic galactic dust-evolution model is presented, where grain growth nicely compensates for the expected higher rate of dust destruction by supernova shocks. This model is simpler, relies on fewer assumptions {\it and} seems to yields a better fit to data derived from observations, compared to previous models of the same type. 
\end{abstract}


\section{Introduction}
The interstellar medium (ISM) - the gas and dust that fills the space between the stars in a galaxy - plays a key role in evolution of a galaxy and, in particular, the build-up and cycling of heavier elements and dust. One of the central problems in cosmic dust evolution is the survival of dust grains. Several decades of research has clearly established \citep[see][any many more]{Barlow78,Draine79,McKee89,Jones96,Slavin04,Jones11} that supernovae (SNe) can induce efficient destruction of dust grains via sputtering by ions associated with the passage of an SN shockwave. The canonical model of SN destruction of dust is due to \citet{McKee89}, which suggest that SNe can effectively cleanse a volume corresponding to an ISM gas mass of order $\sim 1000\,M_{\sun}$ from dust. Recent advances in simulating dust processing have shown that fragmentation \citep[as described in, e.g.,][]{Borkowski95} due to grain-grain collisions can further accelerate the destruction rate, which would lead to efficient cleansing of dust \citep{Kirchschlager19}. But there is yet no actual consensus regarding the dust-destruction rate in the ISM. The high dust masses seen at high redshifts \citep[e.g.,][]{Santini10,Riechers13,Rowlands14a,Rowlands14b,Mattsson15pre,Watson15,Laporte17,Shao19} seem to suggest that cosmic dust forms rapidly and cannot be subject to very efficient destruction processes, since otherwise the net growth of the dust component would be too slow to be consistent with the observations \citep[e.g.,][]{Gall11a,Gall11b,Mattsson11b}. But, at the same time, one has to remember that in the very early Universe (redshifts $z>6$), evidence is mostly lacking of the cold dust seen in the local Universe, except in huge, extreme starburst galaxies (SBGs) with large molecular-gas masses of about $10^{10}\,M_{\sun}$ and estimated star-formation rates (SFRs) of a few times $10^3\,M_{\sun}$\,yr$^{-1}$ . However, in any case, if the fact that large amounts of dust can be formed on very short timescales (at $z \sim 6$ the Universe is just a few hundred Myr) should have to imply low dust destruction rates, that conclusion implicitly assumes that there exists no replenishment mechanism efficient enough to balance the destruction.

Because dust is essentially everywhere in the ISM, cold molecular clouds (MCs) will form with certain amount of dust in them from the beginning. These dust grains may then act as seeds for further dust formations by accretions of molecules of specific types (the ``growth species''), a scenario which has been generally accepted for over 50 years \citep{Baines65b,Baines65c}. In the centres (cores) of cold MCs the number density of molecular gas can often reach $\sim 10^4$cm$^{-3}$ or even more \citep{Sander85,Goldsmith87}, which implies a high probability for accretion of molecules. Interstellar dust formation has been suggested as an important dust-formation channel in many studies, irrespective of redshift and galaxy type \citep[see, e.g.,][]{Draine90,Dwek98,Calura08,Mattsson11b,Valiante11,Asano13a,Ginolfi18}. Direct evidence of this type of grain growth can be difficult to obtain, but there are many indirect indicators of dust grains growing in the ISM. For instance, depletion patterns in interstellar gas are consistent with dust depletion due to grain growth in MCs \citep[see, e.g.,][]{Jenkins09,DeCia16,Mattsson19b}, and late-type galaxies seem to have steeper dust-to-gas gradients than metallicity gradients along the radial extension of their discs, which is easiest explained by grain growth \citep{Mattsson12a,Mattsson12b,Mattsson14b,Vilchez19}.  

In a homogeneous (constant density) environment, grain-growth by accretion is mainly limited by the abundance of the growth-species molecules, which in turn is limited by the overall metallicity (henceforth denoted by $Z$) in the ISM. Thus, the grain-growth timescale is $Z$ dependent. In fact, there even exist a critical $Z$ for efficient grain growth \citep{Asano13a}. Moreover, the growth timescale for an average-sized grain imbedded in a gas of a density corresponding to the average density of an MC, is typically longer than the life time of an MC \citep{Hirashita00}. The growth of dust is in such a case limited by how fast grains grow, rather than how long MCs survive in a state where grains can grow by accretion. However, the ISM in general, and the cold molecular phase in particular, is highly inhomogeneous and display strong gas-density variations on sub-parsec scales. Such gas-density variations mean that some regions have growth-species number densities which are high enough or efficient grain growth.  

Direct numerical simulations of interstellar turbulence \citep[e.g.,][]{Klessen00,Price11,Konstandin12,Federrath13,Nolan15} have established that there exists a direct relationship between the statistical variance of the gas density and the average Mach number (flow speed relative to sound speed). Based on gas-density probability functions derived from a high-resolution compressible hydrodynamics simulations of turbulence in MCs, and the corresponding  relation between Mach number and variance, an accelerated growth rate is expected in supersonic turbulence. The purpose of the present paper is therefore: 
\begin{enumerate}
\item to estimate the effect of gas-density variations due to turbulence on the effective grain-growth velocity; 
\item to explore how such an effect would affect galactic dust evolution. 
\end{enumerate}

\section{Dust growth in the ISM}
\label{dustgrowth}
This section is meant to summarise the elementary theory regarding gas-dust dynamics in a turbulent ISM and dust growth by accretion of metals. A comprehensive summary of galactic dust evolution including interstellar dust growth is also given at the end of the section. 

\subsection{Dynamics and variance of gas and dust}
\label{dynamics}
\subsubsection{Gas}
Interstellar gas is turbulent and highly compressible. Many numerical simulations as well as observational studies suggest root-mean-square Mach numbers $\mathcal{M}_{\rm rms} \gtrsim 10$ \citep[e.g.,][]{Brunt10,Price11,Molina12,Nolan15}, which means the ISM turbulence is clearly in the {\it hypersonic} regime. Consequently, the ISM shows a wide range of gas densities even within cold MCs or the diffuse ISM. Models of interstellar grain growth usually rely on an assumption that the exact gas-density field can be replaced with the mean density, which ``erases'' small-scale variations and other effects of dynamics \citep[but please note the recent examples of inhomogeneous models, e.g.,][although these studies consider variations on much large scales]{Zhukovska16,McKinnon18}. If the density variations are relatively small, this approach is without a doubt very reasonable. But the wide distribution of densities expected due to hypersonic turbulence indicate ``tail effects'', i.e., a significant fraction of the molecular gas in an MC display densities well above the critical density required to obtain growth-species densities high enough to have growth in MC competing with stellar dust production \citep[see][for a more detailed discussion of the critical density]{Asano13a}.

Simulations of isothermal hydrodynamic turbulence with solenoidal (or solenoidally dominated) forcing is known to produce roughly lognormal gas-density statistics \citep[see, e.g.,][and references therein]{Federrath10,Mattsson19a}.  Magneto-hydrodynamic simulations also yield roughly lognormal statistics \citep[see., e.g.,][and references therein]{Molina12}, but with suppressed density variance for very strong turbulence \citep{Ostriker01,Price11}. However, it is the low-density tail that tend to be suppressed \citep{Molina12}, which means that the effect on processes mainly taking place in high-density regions (e.g., dust growth by accretion of molecules) is small. 

The lognormal distribution is of the form
\begin{equation}
\mathcal{P}(s) = {1\over \sqrt{2\pi}\,\sigma_{\rm s}}\exp\left[ -{(s - \mu)^2\over 2\,\sigma_{\rm s}^2}\right], \quad s = \ln\left({\rho\over\langle\rho\rangle}\right), 
\end{equation}
where $\mu$ is related to the variance/standard deviation by $\mu = \langle s\rangle = -{1\over 2}\sigma_{\rm s}^2$ as a consequence of mass conservation \citep{Vazquez94,Konstandin12}, 
\begin{equation}
\int_{-\infty}^\infty \exp({s})\,\mathcal{P}(s)\,ds = \int_0^\infty  \rho\,\mathcal{P}(\rho)\,d\rho  = \langle \rho\rangle,
\end{equation}
which also defines the first moment of $\mathcal{P}(\rho)$. The $n$-th moment of the normalised distribution $\mathcal{P}(\rho)$ is then given by
\begin{equation}
\label{lognormmom}
\left\langle \rho^n\right\rangle = \langle\rho\rangle^n \exp\left[{{1\over 2}\left(n^2 - n\right)\sigma_{\rm s}^2}\right].
\end{equation} 
The variance is given by its relation to the root-mean-square Mach number $\mathcal{M}_{\rm rms}$, usually considered to be of the form
\begin{equation}
\label{sigmamach}
\sigma_{\rm s}^2 = \ln(1+b^2\mathcal{M}_{\rm rms}^2),
\end{equation}
which has been confirmed by several numerical experiments \citep[e.g.,][]{Passot98,Federrath10}. A typical value for the case of for purely solenoidal forcing is $b=1/3$. For mixed forcing, a value $b\approx 0.5$ is often quoted \citep{Federrath13}, which is also the value to be used later.

\subsubsection{Dust}
\label{dustEOM}
Provided that the interstellar dust is only accelerated by interaction with turbulent gas via an \citet{Epstein24} drag law, the Lagranigian equation of motion (EOM) for dust is simple,
\begin{equation}
\label{stokeseq}
{{\rm d} \mathbfit{v}\over {\rm d} t}  = {\mathbfit{u}-\mathbfit{v}\over \tau_{\rm s}},
\end{equation}
where $\mathbfit{v}$ and $\mathbfit{u}$ are the velocities of the dust and the gas, respectively, and $\tau_{\rm s}$ is the stopping time, i.e., the timescale of acceleration (or deceleration) of the grains. In the Epstein limit $\tau_{\rm s}$ depends on the size and density of the grain as well as the gas density and the relative Mach number $\mathcal{W}_{\rm s} = |\mathbfit{u}-\mathbfit{v}|/c_{\rm s}$ \citep{Schaaf63,Baines65}. The dependence on $\mathcal{W}_{\rm s} $ is mathematically complicated, but a simple, yet sufficiently accurate, formula is given by \citep{Kwok75,Draine79}. Non-inertial particles, a.k.a. tracer particles, will have $\mathbfit{v} = \mathbfit{u}$ as well as position coupling with the gas. It is often assumed that the approximation $\mathbfit{v} \approx \mathbfit{u}$ is justified when the stopping time is much shorter than the characteristic timescale of the flow. The latter is a condition  which also depends on spatial resolution. If $|\mathbfit{u}-\mathbfit{v}|\,\tau_{\rm s} \ll \Delta L$, where $\Delta L$ is the spatial resolution, gas and dust can be regarded as effectively position coupled because the flow scale is not resolved.  

An interesting consequence of the fact that $\tau_{\rm s}$ depends on grain size, is that grains of different sizes will have different (number) density distributions, where only the distribution for very small grains agree with the approximately lognormal distribution of the gas \citep{Hopkins16,Mattsson19a}. In particular, there is an anti-correlation between the variance and grain size \citep[see, e.g., Fig. 11 in][]{Mattsson19a}. The critical grain size at which gas and dust start to decouple significantly can be calculated in the low-$\mathcal{M}_{\rm rms}$ limit, because the supersonic correction does not matter that much on average. The stopping time in that limit is $\tau_{\rm s} \sim \rho_{\rm gr}/\langle \rho\rangle\,a/c_{\rm s}$ and an integral timescale of the flow is $\tau_\ell \approx L_\ell/u_{\rm rms}$, where $L_\ell$ is a characteristic length scale in the flow. For an MC we can assume $L_\ell \sim R_{\rm s} \sim 0.1$~pc, where $R_{\rm s} $ is the sonic length \citep[see][]{Hopkins16}. The transition from dynamically coupled to decoupled dust and gas phases happens when $\tau_{\rm s}/\tau_\ell \sim 1$. That is, the transition grain size is given by
\begin{equation}
a_{\rm c} \sim {R_{\rm s}\over\mathcal{M}_{\rm rms}} {\langle \rho\rangle\over \rho_{\rm gr}}.
\end{equation}
For a typical MC, the density ratio $\langle \rho\rangle / \rho_{\rm gr}$ is in the range $10^{-22} \dots10^{-21}$ and with $\mathcal{M}_{\rm rms}= 1\dots 10$ being roughly inverse proportional to  $\langle \rho\rangle$, one can conclude that $a_{\rm c} \sim 1\,\mu$m. If $a \ll a_{\rm c}$ the grains stay coupled to the gas, while if $a \gg a_{\rm c}$ they will decouple and only experience a small drag force from the gas, essentially just a random perturbation. In the present study we will assume that for the vast majority of dust grains in an MC $a \ll a_{\rm c}$. The reasons for this assumption will be explained below.

\subsubsection{Relative velocity}
\label{drift}
The relative velocity (or ``drift velocity''), $\mathbfit{w} = \mathbfit{u}-\mathbfit{v}$, is not necessarily small if the average stopping time is long enough. In such a case, the rate at which gas molecules hit a dust grain cannot be determined only by the abundance and thermal mean speed of relevant molecules. For an individual grain, the relative velocity $\mathbfit{w}$ can be significantly larger than the thermal mean speed. It is easy to show that if all velocity variations can be described as uncorrelated gaussian white noise (which is a reasonable assumption), the root-mean-square value of $\mathbfit{w}$ is given by $w_{\rm rms}^2 = u_{\rm rms}^2  + v_{\rm rms}^2$. If the variance of the velocity distribution for large grains is small compared to that of the gas, then $w_{\rm rms} \approx u_{\rm rms}$, or $\mathcal{W}_{\rm rms} \approx \mathcal{M}_{\rm rms}$. This means that any effect on grain growth that is due to the velocity difference between gas and dust, will also depend on $\mathcal{M}_{\rm rms}$.

\subsection{Moment equations for dust}
\label{momes}
Dust growth by accretion in a multi-dispersed population of dust grains is conveniently described using the method of moments (MOM), which has become the standard method used in models of dust production in stellar atmospheres \citep[see, e.g.,][]{Gail87,Gail88,Ferrarotti06,Mattsson10a,Mattsson11a,Ventura12}, but can be adapted to interstellar dust processing as well \citep{Mattsson16}. However, as shown below, the method may become somewhat inconvenient when dust--gas interaction is taken into account. But the difficulties will vanish if the method of moments is applied to an ensemble of grains, which can be considered in terms of spatial averages as in kinetic theory. 

The moments of order $\ell$ of the grain-size distribution (GSD) $f(a,t)$ are defined as
\begin{equation}
\mathcal{K}_\ell(t) = \int_0^\infty a^\ell f(a,t)\,da
\end{equation}
where $a$ is the grain radius of a spherical grain. In a static medium, the system of differential equations governing the hierarchy of moments of the GSD under condensational growth is given by
\begin{equation}
{d\mathcal{K}_\ell\over dt} = \ell \xi(\mathbfit{x}, t)\,\mathcal{K}_{\ell-1}(\mathbfit{x}, t),
\end{equation}
where $\xi(\mathbfit{x},t)=da/dt$ is the thermal grain-growth velocity, i.e., the rate at which $a$ increases due to thermal collisions (and chemical reactions) with the considered molecular growth species.$\mathbfit{x}$ is the spatial position vector. The corresponding mass growth of a grain of radius $a$ can be expressed as \citep{Hirashita11,Mattsson14a}
\begin{equation}
\label{massgrowth}
{dm_{\rm gr}\over dt} =  {4\pi}\,a^2 S\,X_i\,\bar{u}_{\rm t} \,\rho(\mathbfit{x},t)
\end{equation}
where $X_i$ is the mass fraction of the relevant growth-species molecules $i$ in the gas, $S$ is the sticking probability for a molecule hitting the grain and $\bar{u}_{\rm t}$ is the thermal mean speed of the molecules (which is assumed to be constant). The mass of a spherical grain is $m_{\rm gr} = 4\pi/3\,\rho_{\rm gr}\,a^3$, where $\rho_{\rm gr}$ is the bulk material density. Taking the derivative with respect to $a$ yields 
\begin{equation}
{dm_{\rm gr}\over da} = 4\pi\,\rho_{\rm gr}\,a^2, 
\end{equation}
and division by equation (\ref{massgrowth}) leads to an equation for the growth velocity
\begin{equation}
\xi(\mathbfit{x},t) = S\,\bar{u}_{\rm t}\,X_i(t)\,{\rho(\mathbfit{x},t)\over  \rho_{\rm gr}},
\end{equation}
which is independent of the grain radius $a$. If the molecular composition of the gas remains well mixed, so that $X_i$ only depends on time, $\xi$ can be conveniently written as
\begin{equation}
\label{growthvel}
\xi(\mathbfit{x},t)  = \langle \xi\rangle\,{\rho(\mathbfit{x},t)\over\langle\rho\rangle},\quad  \langle \xi\rangle =S\,\bar{u}_{\rm t}\,X_i(t)\,{\langle\rho\rangle\over  \rho_{\rm gr}},
\end{equation}
where $\langle\rho\rangle$ is the mean density of the gaseous medium in consideration.

\subsection{Spatial-mean equations}
In order to include the effects of a turbulent, highly compressible ISM, the moment equations must be combined with the EOM for the dust component. This results in a complicated system of equations, whose solution requires a numerical approach that is computationally expensive. Direct numerical simulation of turbulence including dust growth by accretion via inertial ``super particles'' \citep[see][for an example regarding coagulation]{Zsom08} is likely a better approach, but such simulations are also computationally demanding. Thus, to test the hypothesis that high-$\mathcal{M}_{\rm rms}$ turbulence can accelerate dust growth, described in Section \ref{momes} above, it is therefore very reasonable to start with some kind of spatial-mean approach which may allow use of the conventional MOM. 

\subsubsection{General case}
If the velocities of the dust grains are taken into account, the moment equations must include a term describing the advection of dust, 
\begin{equation}
\label{momflow}
{\partial \mathcal{K}_\ell \over \partial t} = \ell\,\xi(\mathbfit{x}, t)\,\mathcal{K}_{\ell - 1}(\mathbfit{x}, t) + \int_0^\infty a^\ell f(a,\mathbfit{x}, t)\,\nabla\cdot \mathbfit{v}(a, \mathbfit{x},t)\,da.
\end{equation}
The last (integral) term arises from the fact that the velocities of the grains depend on the sizes of the grains (see equation \ref{stokeseq}). In case of exact velocity coupling between gas and dust, i.e., if the dust grains behave as tracer particles, the moment equations can be solved as part of the flow problem. But as shown by, e.g., \citet{Hopkins16,Mattsson19a}, dust and gas in a turbulent molecular cloud can show vastly different dynamical behaviour. Unless the condensation problem for a sufficiently large number of grain sizes (bins) is solved together with the hydrodynamic flow, the integral term on the right-hand-side of equation (\ref{momflow}) remains undetermined. A different approach is clearly needed. 

A spatial mean, taken over a large enough volume, will solve the problem in the sense that the integral term may be assumed to vanish. The form of the equation  is then the same as when a Lagrangian frame is adopted. One can also say that this spatial mean is a {\it Lagrangian mean} when $\langle \mathbfit{v}\rangle = 0$ \citep{Holm99}. Thus, this type of mean is sometimes referred as a Lagrangian mean, although it may be technically incorrect. In the following the term ``spatial mean'' will be used.

The spatial-mean approach is very useful for periodic-box models because the divergence of any vector field {\it must} have a vanishing mean in a periodic box, which is why the integral term in equation (\ref{momflow}) will make no net contribution. Thus, the mean moment equations are 
\begin{equation}
\label{generic}
{d\langle \mathcal{K}_\ell\rangle\over dt} \approx \ell\, \langle\xi \, \mathcal{K}_{\ell-1}\rangle,
\end{equation}
which is essentially just the mean of the usual set of equations. Unfortunately, these equations cannot be solved as a moment hierarchy, because the mean on the right-hand-side is taken over a product of $\xi$ and $\mathcal{K}_{\ell-1}$. Two limiting cases can be considered as approximations, however, which will be described below.


\subsubsection{Tracer-particle limit} 
In many astrophysical flows, dust particles have relatively short stopping times and must couple well to the gas flow on the scales that can be observed. In such a case, effects due to the relative velocity $\mathbfit{w}$ between gas and dust are negligible and the dust-to-gas ratio $\psi = n_{\rm d}/n_{\rm mol}$, where $n_{\rm d} = \mathcal{K}_0$ is the number density of dust grains and $n_{\rm mol}$ that of gas particles/molecules, remains constant with respect to space and time if the initial condition is spatially invariant (i.e., the gas and the dust are initially well mixed). The moments can be expressed
\begin{equation}
\mathcal{K}_\ell  =  {\mathcal{K}_\ell \over \mathcal{K}_0} \,n_{\rm d} = \psi_0\,{\mathcal{K}_\ell \over \mathcal{K}_0} \,{\rho \over m_{\rm mol}},
\end{equation}
where $m_{\rm mol}$ is the mean molecular mass for the interstellar gas and $\psi_0$ is the initial/average $\psi$, which remains constant when dust and gas are coupled\footnote{The dust number density $n_{\rm d}$ is obviously constant on average if shattering and coagulation/aggregation is not considered. $\langle\rho\rangle$, on the other hand, will be affected by the phase transition taking place when molecules hit and react with dust grains. But this depletion of molecules is so small that $\langle\rho\rangle$ can be regarded as a constant to first approximation. Thus, $\psi_0/m_{\rm mol} = \langle n_{\rm d}\rangle/\langle\rho\rangle =$~constant.}. Thus, starting from eq. (\ref{generic}) and recalling that $\xi(\mathbfit{x},t)/\langle \xi\rangle = \rho(\mathbfit{x},t)/\langle\rho\rangle$, it is straight forward to show that the spatial-mean moment equations for the tracer-particle limit can be written
\begin{equation}
\label{tracermoments}
{d\langle \mathcal{K}_\ell\rangle\over dt}  = \ell\, {\langle\xi\rangle\over \langle\rho\rangle}\,\left\langle {\psi_0\over m_{\rm mol}}{\mathcal{K}_{\ell-1}\over \mathcal{K}_0} \,\rho^2 \right\rangle.
\end{equation}
Since $\psi_0/m_{\rm mol} = \langle n_{\rm d}\rangle/\langle\rho\rangle$, the particular case  $\ell = 1$  leads to an equation for the average grain radius $\langle a\rangle \equiv \langle \mathcal{K}_1 \rangle /  \langle \mathcal{K}_0 \rangle = \langle \mathcal{K}_1 \rangle /  \langle n_{\rm d} \rangle$,
%
\begin{equation}
\label{atracer}
{d\langle a\rangle\over dt}  
=  \langle \xi\rangle \,{\langle \rho^2 \rangle \over \langle \rho \rangle^2},
\end{equation}
where the brackets around $a$ symbolises both a spatial mean and a grain-population average at the same time.
Combining Eqs. (\ref{lognormmom}), (\ref{sigmamach}) and (\ref{atracer}) one can then write 
\begin{equation}
\label{tracerapprox}
{d\langle a \rangle\over dt} = \langle \xi \rangle \, (1+b^2\mathcal{M}_{\rm rms}^2),
\end{equation}
which implies that the effective average growth velocity increases rapidly with increasing $\mathcal{M}_{\rm rms}$ and  $\langle \xi \rangle = \langle da/dt \rangle$ only when $\mathcal{M}_{\rm rms}~\to~0$. One can also conclude, from numerical tests, that the moment equations are reasonably well-approximated with
\begin{equation}
\label{tracermoments_approx}
{d\langle \mathcal{K}_\ell \rangle\over dt} \approx \ell\langle \xi \rangle\,\langle \mathcal{K}_{\ell-1}\rangle \, (1+b^2\mathcal{M}_{\rm rms}^2),
\end{equation}
which will be used later.

\subsubsection{Large-particle limit}
Grains with large enough inertia and long stopping times (grains with radii $a\gtrsim 1\,\mu$m as shown in in Section \ref{dustEOM}) justify the assumption that the velocity distributions for gas and dust are statistically independent, defines another important limit. Actually, there are two limits to consider here: that which is obtained for low mean Mach numbers, i.e.,  $\mathcal{M}_{\rm rms}\ll 1$, and that which is obtained for  $\mathcal{M}_{\rm rms}\gg 1$. 

In the first case, where the effects of the relative velocity $\mathbfit{w}$ are small, and dust grains and gas are uncorrelated, it is fair to assume $\langle\xi\,\mathcal{K}_{\ell-1}\rangle \approx \langle\xi\rangle\,\langle\mathcal{K}_{\ell-1}\rangle$. In such a case, the mean moment equations simply become,
\begin{equation}
\label{largeapprox1}
{d\langle \mathcal{K}_\ell\rangle\over dt} \approx \ell\, \langle\xi\rangle\,\langle \mathcal{K}_{\ell-1}\rangle,
\end{equation}
which is, of course, the same as equation (\ref{tracerapprox}) in the weakly compressible regime\footnote{It is worth mentioning that for very small Mach numbers $\mathcal{M}_{\rm rms}$, the gas density should remain essentially uniform if the initial distribution was uniform ($\sigma_{\rm s}\to 0$), which is the same as to say that the dust-condensation problem follow the usual spatially independent formulation. A uniform gas distribution is, however, not realistic in an astrophysical context.} ($\mathcal{M}_{\rm rms}\ll 1$).

For $\mathcal{M}_{\rm rms}\gg 1$, the decoupling between gas and dust becomes important. \citet{Baines65} showed that if $\mathbfit{w}\gg \bar{u}_{\rm th}$, then, to first order,
\begin{equation}
{da\over dt} \approx {\xi_0\over 4}{|\mathbfit{w}|\over \bar{u}_{\rm th}}{\rho\over \rho_0},
\end{equation}
where the factor of four in the denominator comes from the fact that accretion onto a rapidly moving grain is limited by its cross-section rather than its total surface area. Under the assumption that $ \bar{u}_{\rm th}$ is Maxwellian, $\xi$ must be proportional to the relative Mach number $\mathcal{W} = |\mathbfit{w}|/c_{\rm s}$, which means that the mean growth velocity is
\begin{equation}
\left\langle {d a\over dt} \right\rangle \approx \sqrt{\pi\over 128}{\langle\xi\rangle\over \rho_0}\langle \mathcal{W}\,\rho\rangle.
\end{equation}
The mean gas density $\langle \rho\rangle = \rho_0$ and relative Mach number $\langle \mathcal {W}\rangle$ can be assumed to arise from independent (disjoint) distributions. Thus, $\langle \mathcal{W}\,\rho\rangle \approx \langle \mathcal{W}\rangle \,\rho_0$. Furthermore, $\langle \mathcal {W}\rangle \approx \sqrt{8/3\pi}\,\mathcal{W}_{\rm rms} $ if $|\mathbfit{w}|$ has a Maxwellian distribution and $\mathcal{W}_{\rm rms} = w_{\rm rms}/c_{\rm s} \approx \mathcal{M}_{\rm rms}$ (see Section \ref{drift}), which yield approximate moment equations of the form
\begin{equation}
\label{largeapprox2}
{d\langle \mathcal{K}_\ell\rangle\over dt} \approx {\mathcal{M}_{\rm rms}\over 4\sqrt{3}}\, \ell\,\langle\xi\rangle\,\langle \mathcal{K}_{\ell-1}\rangle.
\end{equation}
From the equations above it is obvious that if $\mathcal{M}_{\rm rms}\sim 10$ the turbulence effect is still not very significant. 

On a more general note, the fact that gas-density fluctuations do dot seem to have any net effect in the large-particle limit can be understood as a reflection of the linear dependence on the gas density ($\xi\propto \rho$). Increased growth efficiency in high-density regions seems to be compensated by lower efficiencies in the low-density voids; on average there is neither an increase, nor a decrease of the rate.

\section{Implications for galactic dust evolution}

Once this new formulation of the dust-growth rate in MCs has been obtained, it could be of interest to consider a simple model of galactic dust evolution with interstellar dust growth. The present section aims to derive the grand-scale implications of the theory in section \ref{dustgrowth}, after first formulating a  simple model of galactic dust evolution (GDE).

\subsection{Formulation of the GDE model}
\subsubsection{Simplifying assumptions} 
To minimise the number of free parameters one can make a few simplifying assumptions. First, the the stellar dust/metals production can be described under the instantaneous recycling approximation (IRA), i.e., the lifetimes of stars are negligible compared to overall evolutionary timescale. Second, one may also assume that the effects of the evolution of the GSD are small on average, so that grain growth and destruction are functions of macroscopic properties only as described in the next section. Third, the fraction of metals available for accretion (metals that may end up in dust) $\tilde{Z}$ is essentially the same as the mass fraction of metals not (yet) locked up in dust, i.e.,   $\tilde{Z} \approx Z-Z_{\rm d}$, where $Z$ and $Z_{\rm d}$ are the total mass fractions of metals and dust, respectively. The latter assumption is reasonable because the observed depletion is surprisingly close to 100\% for many of the most abundant metals except C, N, O and noble gases \citep[see, e.g.,][]{Pinto13,DeCia16}. 

Consider now a system of total mass $M = M_{\rm s} + M_{\rm g}$, where $M_{\rm s}$ and $M_{\rm g}$ are the masses of stars and gas, respectively. The system is assumed to be a ``closed box'', which corresponds to $dM/dt = 0$ and can be seen as the limit in which galaxy formation is much faster than the build-up of heavier elements. This assumption greatly simplifies the model and, together with the previous assumptins above, it will also allow exact solution of the dust-evolution equation. Outflows (``galactic winds'') can alter the evolution too, but mainly by altering the effective yields (see results, Section \ref{outflows}). 

\subsubsection{Dust-evolution equation}
With the assumptions above, the equation for the dust-to-gas ratio $Z_{\rm d}$ can be written \citep{Mattsson12a},
\begin{equation}
\label{dtgeq}
{dZ_{\rm d}\over dZ} =  {y_{\rm d}\over y_Z} + {Z_{\rm d}\over y_Z}[G(Z)-D(Z)] ,
\end{equation}
where $G$ is the rate of increase of the dust mass due to grain growth relative to the rate of gas consumption due to star formation, $D$ is the corresponding function for dust destruction and $y_{\rm d}$, $y_Z$ are the effective stellar dust and metal yields, respectively. Both $y_{\rm d}$ and $y_Z$ may depend on the $Z$, but will be regarded as constants to first approximation.

It has been argued in previous works \citep{Mattsson12a,Mattsson12b,Mattsson14a,Mattsson14b,Rowlands14a,Rowlands14b} that dust growth would be the most important mechanism for changing the dust-to-metals ratio $\zeta = Z_{\rm d}/Z$ in a galaxy throughout its course of evolution as well as creating a dust-to-metals gradient along galaxy discs. Hence, it can be worthwhile writing down an equation for $\zeta$ as well,
\begin{equation}
\label{dtmeq}
Z{d\zeta\over dZ} =  {y_{\rm d}\over y_Z} + {Z \zeta \over y_Z}[G(Z)-D(Z)]- \zeta,
\end{equation}
where all other quantities are as previously defined. 
Explicit forms of the functions $D$ and $G$ to be used for modelling will be discussed in Section \ref{results}.

\subsection{Implications of turbulence accelerated grain growth}

In order to formulate an explicit functional form of $G$, one must compare the different timescales involved in the dust growth in the MC. There are, essentially, three different timescales here: the MC formation time $\tau_{\rm form}$, which defines the MC formation rate; the growth timescale of the dust within an MC $\tau_{\rm grow}$; and the characteristic lifetime of MCs $\tau_{\rm MC}$.  However, there is also a fourth timescale, the cycling time $\tau_{\rm cyc}$, i.e., how long it would typically take for an atom in the diffuse ISM to cycle through the cold phase and being returned to the diffuse ISM again. $\tau_{\rm cyc}$ is roughly the sum $\tau_{\rm form}$ and $\tau_{\rm MC}$ and all three timescales are of the same order of magnitude, which will be described below.

\subsubsection{Limit cases}
The galactic-scale evolution of the MC phase, neglecting the effects of star formation, is governed by an equation of the form
\begin{equation}
{dM_{\rm MC}\over dt} \eqsim \left({1\over \eta}-1\right) {M_{\rm MC}\over \tau_{\rm form}}- {M_{\rm MC}\over \tau_{\rm MC}},
\end{equation}
where $\eta = M_{\rm MC}/M_{\rm g}$. According to \citet{Elmegreen90}, $\tau_{\rm form} \sim \tau_{\rm MC}$, which can also be understood by assuming an equilibrium state, i.e., ${dM_{\rm MC}/ dt}=0$. In combination with the fact that gas-mass estimates of late-type galaxies imply $\eta \sim 1/2$, this equilibrium suggests that $\tau_{\rm MC} \sim \tau_{\rm form}$. That is,  $\tau_{\rm cyc}$, $\tau_{\rm form}$ and $\tau_{\rm MC}$ are of the same order of magnitude. More generally, however, the equilibrium $\eta$ is not always $\sim 1/2$ and $\tau_{\rm MC} \sim \tau_{\rm form}$ may not hold. If $\eta < 1/2$, then $\tau_{\rm form} > \tau_{\rm MC}$ and, vice versa, if  $\eta > 1/2$, then $\tau_{\rm form} < \tau_{\rm MC}$. 

Gas is converted into stars at a rate $dM_{\rm s}/dt \approx M_{\rm MC}/\tau_{\rm sfr}$, where $\tau_{\rm sfr}$ is the star-formation timescale. This rate is mainly regulated by $M_{\rm MC}$ (stars form from the cold phase), so that $dM_{\rm s}/dt \approx dM_{\rm MC}/dt$.  Hence,
\begin{equation}
{dM_{\rm s}\over dt} \propto {M_{\rm MC}\over \tau_{\rm form}}\sim {M_{\rm MC}\over \tau_{\rm MC}},
\end{equation}
or, the timescales are related as $\tau_{\rm sfr}\propto \tau_{\rm form}\sim \tau_{\rm MC}$.

In GDE models, it is relevant how $\tau_{\rm grow}$ compares to $\tau_{\rm cyc}$ or $\tau_{\rm MC}$. A simple estimate yields that $\tau_{\rm grow}> \tau_{\rm MC}$ in a typical MC, assuming a homogeneous distribution of matter (see example in Section \ref{growthinMCs}). In case $\tau_{\rm grow} \gg \tau_{\rm MC}$ (grain growth clearly not accelerated by turbulence), it is $\tau_{\rm grow}$ that limits the overall rate of growth. The growth of the dust density $\rho_{\rm d}$ in MCs follows the third moment $\langle K_3\rangle$. Assuming $\langle K_3\rangle \approx {1\over 3}\langle K_2\rangle/a_0$, where $a_0 = \langle a(0) \rangle$ is the initial mean grain radius, one can show that \citep{Mattsson12a,Mattsson16},
\begin{equation}
{1\over \tau_{\rm grow}} \approx {1\over \tau_0(Z)}\left(1 - {Z_{\rm d}\over Z}\right)
\end{equation}
where $\tau_{0} \equiv a_0/\xi_0$, $\xi_0=S\,\bar{u}_{\rm t}\,X_i(0)\,\langle\rho\rangle/ \rho_{\rm gr}$ are the initial growth timescale and mean growth velocity, respectively. Because $\tau_{\rm grow} \gg \tau_{\rm MC}\sim \tau_{\rm cyc}$, any significant dust growth must occur via several cycles in and out of the cold phase, which means that $\tau_0$ will effectively be inverse proportional to $\eta\,Z\,M_{\rm g}$, where $\eta\,M_{\rm g} = M_{\rm MC} \propto dM_{\rm s}/dt$. Thus, the growth model used by \citet{Mattsson14a} is obtained, i.e.,
\begin{equation}
\label{growthrate2}
M_{\rm g}\,\left( {dZ_{\rm d}\over dt} \right)_{\rm grow} \propto  Z_{\rm d} \left(Z - {Z_{\rm d}}\right)\,{dM_{\rm s}\over dt}.
\end{equation}

In the opposite limit,  $\tau_{\rm grow}\ll \tau_{\rm MC}$, i.e., in the case of accelerated grain growth, the rate of dust growth is determined by the formation timescale $\tau_{\rm form}$. That is, one may view MCs as dust producers in the same way as stars: they are forming at a certain rate (defined by $\tau_{\rm form}$); exist for a limited time, which is short compared to the galaxy-evolution timescale and can therefore be ignored (a type of IRA for MCs); the amount of dust formed depends on the amount of gas-phase metals available -- not the growth timescale $\tau_{\rm grow}$ -- a fact which can be treated as if there is a ``yield'' for MCs, $y_{\rm MC}\propto Z-Z_{\rm d}$. Thus,
\begin{equation}
\left({dM_{\rm d}\over dt}\right)_{\rm grow} \sim (Z-Z_{\rm d})\, {M_{\rm g}\over \tau_{\rm form}}. 
\end{equation}
Under the assumption of some kind of IRA for MCs as described above, the dust-growth rate is again proportional to the star formation rate ($\tau_{\rm sfr} \propto \tau_{\rm form}$), so that
\begin{equation}
\label{growthrate1}
M_{\rm g}\,\left( {dZ_{\rm d}\over dt} \right)_{\rm grow} \propto (Z-Z_{\rm d})\,{dM_{\rm s}\over dt}.
\end{equation}
The connection between the star-formation rate and the dust-growth rate in MCs is indeed not a  new idea \citep[see, e.g.,][]{Hirashita11}. The difference is that in turbulence-accelerated growth, the $Z_{\rm d}^2$ term does not appear in eq. (\ref{growthrate1})  as in eq. (\ref{growthrate2}), which is an indirect consequence of the short $\tau_{\rm grow}$. In section \ref{growthinMCs} below the shortening of the timescale $\tau_{\rm grow}$ will be demonstrated in a more quantitative way.

\section{Results and discussion}
\label{results}
\subsection{Dust growth rate in turbulent MCs}
\label{growthinMCs}
It is well established that dust and gas inside an MC can show significant dynamical decoupling (drift). But on the scale of an entire MC, the vast majority of dust grains can be regarded as spatially coupled to the gas. Because the GSD falls steeply with grain radius and the rate of accretion of metals onto pre-existing dust is largely determined by the total grain-surface area, it is trivial to show that most of the dust-mass growth is due to small dust grains \citep[see, e.g., Fig. 3 in][]{Hirashita11}. Using the tracer-particle limit to model the growth rate is therefore justified in most cases. 

\begin{figure}
  \resizebox{\hsize}{!}{
   \includegraphics{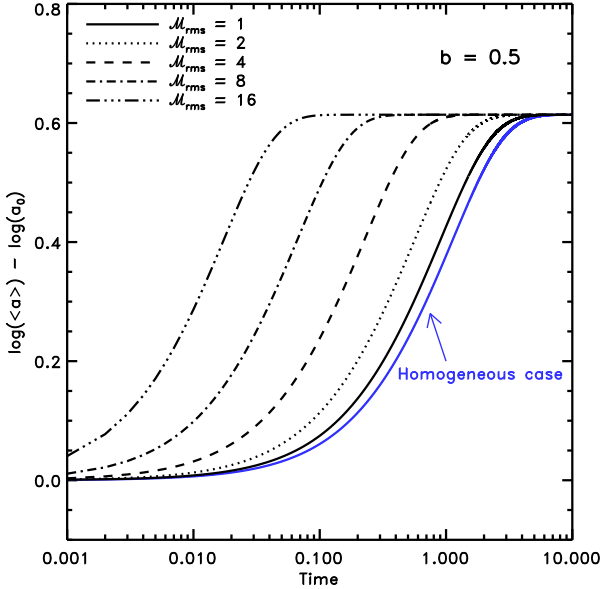}
   }
  \caption{\label{a_evol} {Time evolution of the average grain radius  $\langle a\rangle \equiv \langle \mathcal{K}_1 \rangle /  \langle \mathcal{K}_0 \rangle$ for different root-mean-square Mach numbers $\mathcal{M}_{\rm rms}$. The time unit is the initial growth timescale for the homogeneous case, defined as $\tau_{\rm grow} = a_0/\langle \xi \rangle$, where $a_0$ is the initial mean radius.}}
  \end{figure}

By numerically solving for (at least) the first four moments defined by equation (\ref{tracermoments_approx}),  a good estimate of the growth of the mean grain radius $\langle a\rangle$ is obtained. Provided that just one generic dust species is considered and there is no injection of the growth-species molecule(s) into the system, one may write $X_i(t) = X_i(0) - \langle\rho_{\rm d}\rangle/\langle\rho\rangle$, where $\langle\rho_{\rm d}\rangle = 4\upi/3\,\rho_{\rm gr}\langle\mathcal{K}_3\rangle$. The expression for $\langle\xi\rangle$ then becomes
\begin{equation}
\langle\xi\rangle = S\,\bar{u}_{\rm t}\,\left[X_i(0) {\langle\rho\rangle\over \rho_{\rm gr}} -  {4\upi\over 3}{\langle \mathcal{K}_3\rangle} \right].
\end{equation}
The system of equations (\ref{tracermoments_approx}), including the equation above, is closed and can easily be solved using a modified module of the MOMIC code \citep{Mattsson16}, which gives the sigmoid-type solutions for $\langle a \rangle$ seen in Fig. \ref{a_evol}. The time unit in Fig. \ref{a_evol} is the initial growth timescale for the homogeneous case $\tau_0$, as previously defined.  Clearly, dust growing in an MC with $\mathcal{M}_{\rm rms}\sim 10$ will reach the saturation limit much faster than dust growing in a non-turbulent gas with only small density variations. Assuming the initial GSD is the canonical MRN distribution \citep{Mathis77}, the mean radius is of the order $a_0\sim 0.01\,\mu$m. Then, assuming a molecular number density $n_{\rm mol} \sim 100$\,cm$^{-3}$, a thermal mean speed $\bar{u}_{\rm t} = 0.15$\,km\,s$^{-1}$ (corresponding to $T_{\rm gas} = 40-50$\,K),  maximum sticking probability ($S=1$) and a grain-material density $\rho_{\rm gr} = 2.4$\,g\,cm$^{-3}$, equation (\ref{growthvel}) suggest $\tau_{\rm grow} \sim 10^8$\,yr (in some MCs $T_{\rm gas}$ is lower and $n_{\rm mol}$ higher, but $\tau_{\rm grow}$ is of the same order of magnitude). This is longer than the theoretically expected lifetime of an MC \citep[$\tau_{\rm MC} \sim 10^7$ yr, see, e.g.,][]{Elmegreen90}, which would suggest that dust depletion in MCs is limited by their disruption. It should be noted, though, that average MCs can locally have $n_{\rm mol} \sim 10^5$\,cm$^{-3}$  (or even higher), which implies that in some dense regions of an MC $\tau_{\rm grow} < \tau_{\rm MC}$. But for such an MC as a whole, the typical density is rather of the order $n_{\rm mol} \sim 100$\,cm$^{-3}$  \citep{Sander85}.

According to Fig. \ref{a_evol}, the effective $\tau_{\rm grow}$ can be reduced by to two orders of magnitude, in which case the grain growth may saturate within the lifetime of an MC, i.e., the metals in an MC can in fact be almost fully depleted. That is, the dust depletion is {\it not} limited by the timescale ratio $\tau_{\rm grow}$. It would in such a case rather be controlled by the MC formation timescale, which is also expected to be of order $10^7$\,yr \citep[again, see][]{Elmegreen90}. Since $10^7$\,yr is a short time compared to the overall evolutionary timescale of cosmic dust and metals,  the grain-growth rate is more or less directly proportional to the formation rate of MCs. This reduction reflects that high-density regions have much shorter local $\tau_{\rm grow}$, as mentioned above, which combined with the fact that most of the gas mass is found in dense clumps explain why taking an inhomogeneous gas distribution due to high-$\mathcal{M}_{\rm rms}$ turbulence into account will  lead to $\tau_{\rm grow} < \tau_{\rm MC}$ also for the MC as a whole.

In summary: the effective dust-growth timescale $\tau_{\rm grow}$ in a highly turbulent MC is clearly much shorter than in a homogeneous MC; assuming typical scalings, the lifetime of an MC is much longer than $\tau_{\rm grow}$; the lifetimes of MCs are in turn short compared to the overall evolutionary timescale of interstellar dust. Consequently, dust production by accretion of metals in MCs in a galaxy is essentially regulated by the formation rate of MCs.

\subsection{Galactic dust evolution with rapid growth in MCs}
\subsubsection{Stellar dust and metals production and galactic outflows}
\label{outflows}
The stellar yields are obtained by summing up the ejecta of newly produced metals in total ($y_Z$), or the fraction which is in the form of dust ($y_{\rm d}$), for a generation of stars \citep[see][for a definition of stellar yield]{Pagel97}. This means that $y_{\rm d}$ and $y_Z$ are directly dependent on the shape of the stellar initial mass function (IMF) and stellar $Z$. Although the IMF is not completely invariant from one local environment to another, it is still surprisingly invariant on average \citep{Bastian10}. Significant IMF evolution is likely occurring at very early stages and may affect the evolution of certain abundance ratios for very unevolved systems, but does not seem to be important at later stages \citep[see, e.g.,][]{Chiappini00,Mattsson10b}. Similarly, the composition of the ejecta of newly formed elements from evolved stars must in principle depend on the stellar $Z$, but this dependence is strong only at very low $Z$ \citep[see, e.g., the discussion about yields in][]{Mattsson10b}. Thus, $y_{\rm d}$ and $y_Z$ can, to first approximation, be regarded as constants throughout the course of evolution of most galaxies.

Despite the fact that approximately constant yields appear reasonable, there is one other mechanism one has to take into account here: galactic outflows/winds. Especially low-mass star-bursting galaxies (a prototypical local example is I Zw 18) may have gaseous outflows due to radiation pressure and kinetic-energy injection by SNe. The former, in particular, may have a strong connection to dust; a dust-driven galactic wind can be formed in much the same way as dust-driven stellar winds \citep{Nath09,Thompson15,Costa18}. Because dust and gas are not perfectly coupled (when dust grains are large) and do not behave as a single ``fluid'' (see section \ref{dustEOM}), the loss of dust may be relatively higher than the loss of metals due to such an outflow.\footnote{Note, however, that it can as well be argued that the dust-to-metals ratio is suppressed by non-selective outflows in combination with enriched inflows \citep{Feldmann15}. Also in such a case, the effect can be seen as a lowering of the effective dust yield.} The simplest model of the galactic outflow rate would be direct scaling with the star-formation rate, which can be motivated by the connection between star-formation rate and the strength of the radiation field in a galaxy as well as the kinetic- and thermal-energy input from stars (mainly SNe). 
The conventional parameterisation would be \citep{Matteucci83}
\begin{equation}
\left( {dM_{\rm g}\over dt} \right)_{\rm out} = -w\,{dM_{\rm s}\over dt},
\end{equation}
where $w$ is an efficiency factor which is either describing the momentum transfer from radiation via dust grains or heating and dissipation of kinetic energy from SNe (or a combination of both). In either case, the net effect is the same as altering $y_{\rm d}$ and $y_Z$ \citep{Avila16}, or more precisely, 
\begin{equation}
{dZ_{\rm d}\over dZ} = {y_{\rm d} (1+w_{\rm g})\over y_Z (1+w_{\rm d})} +  {Z_{\rm d}\over y_Z (1+w_{\rm d})}[G(Z)-D(Z)] ,
\end{equation}
where the factors $w_{\rm g}$ and $w_{\rm d}$ corresponds to the ``wind efficiency'' for gas and dust, respectively. This suggests there is a wide range of effective values of  $y_{\rm d}$ and $y_Z$ applying to various different systems and environments. The effect of varying $y_{\rm d}$ and $y_Z$ is mostly seen at early stages and seems to explain the large statistical variance in observed dust-to-metals ratios (see Fig. \ref{dtm_dep}). 

\subsubsection{Grain growth}
The result described above favour a model of grain growth, which is based on the MC formation rate. 
From eq. (\ref{growthrate1}) it follows that
\begin{equation}
\left( {dZ_{\rm d}\over dZ} \right)_{\rm grow} \propto {1\over y_Z}\,(Z-Z_{\rm d}).
\end{equation}
Hence, the rate of increase of $M_{\rm d}$ due to grain growth relative to the rate of gas consumption due to star formation, $G$, can be written,
\begin{equation}
G(Z) = {\epsilon}\,\left[{Z\over Z_{\rm d}(Z)}-1\right], 
\end{equation}
where  $\epsilon$ is a generic efficiency factor. This factor, $\epsilon$, is proportional to $\mathcal{M}_{\rm rms}^2$, but direct parameterisation in terms of  $\mathcal{M}_{\rm rms}$ will be degenerate and is therefore not meaningful.

\subsubsection{Dust destruction by SNe}
The dominant dust-destruction mechanism is sputtering in the high-velocity interstellar shocks driven by SNe, which can be directly related to the energy of the SNe \citep{Nozawa06}.
Following \citet{McKee89,Dwek07} the dust destruction time-scale is
\begin{equation}
\tau_{\rm d} = {\rho\over \langle m_{\rm ISM}\rangle\,\mathcal{R}_{\rm SN}},
\end{equation}
where $\rho$ is the gas mass density, $\langle m_{\rm ISM}\rangle$ is the effective gas mass cleared of dust by each SN event, and $R_{\rm SN}$ is the SN rate per volume. Due to the short evolutionary timescale of massive stars, the latter is
\begin{equation}
\label{snr}
\mathcal{R}_{\rm SN}(t) \approx \mathcal{R}_{\rm sfr}(t)\int_{8M_\odot}^{100M_\odot} \phi(m)\,dm,
\end{equation}
where $\phi(m)$ is the stellar IMF and $\mathcal{R}_{\rm sfr}$ is the star-formation rate per unit volume. For a universal IMF the integral in equation (\ref{snr}) is a constant with respect to time, and space. Hence, 
\begin{equation}
\label{taud}
\tau_{\rm d}^{-1} \approx  {\delta\over M_{\rm g}}{dM_{\rm s}\over dt} = {\delta\over y_Z}{dZ\over dt},
\end{equation}
where $\delta$ is an arbitrary dust-destruction efficiency parameter. The destruction rate relative to the rate of gas consumption $D$ is then simply $D = \delta$.

\subsubsection{Exact solution}
With constant stellar yields and the prescriptions for $G$ and $D$ described above, equation (\ref{dtgeq}) takes the specific form
\begin{equation}
{dZ_{\rm d}\over dZ} = {y_{\rm d}\over y_Z} + {Z_{\rm d}\over y_Z}\left[{\epsilon} \left({Z\over Z_{\rm d}} - 1\right) - \delta\right],
\end{equation}
and, similarly, equation (\ref{dtmeq}) takes the form
\begin{equation}
Z{d\zeta\over dZ} = {y_{\rm d}\over y_Z} + {Z\zeta\over y_Z}\left[{\epsilon} \left({1\over \zeta} - 1\right) - \delta\right] -\zeta.
\end{equation}
These equations are solved by
\begin{equation}
\label{solution}
Z_{\rm d} = Z\zeta = {\epsilon\over \delta + \epsilon}\left\{\left({y_{\rm d}\over \epsilon} - {y_Z\over \delta + \epsilon}\right)\left[1-\exp\left(-{{\delta+\epsilon\over y_Z}\, Z}\right) \right] + Z\right\},
\end{equation}
which assume an initial condition $Z_{\rm d}(0) = 0$.

\subsubsection{The special case $\delta = \epsilon$}
\begin{figure}
  \resizebox{\hsize}{!}{
   \includegraphics{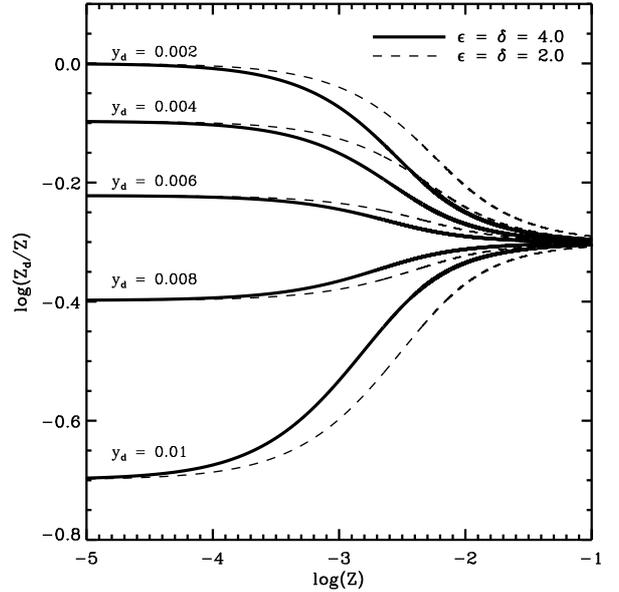}
   }
  \caption{\label{var_yd} Dust-to-metals ratio as a function of metallicy. Variation of the stellar dust yield $y_{\rm d}$ for two cases where $\epsilon = \delta$ (see Section \ref{compobs} for further details).}
  \end{figure}

The equilibrium model suggested by \citet{Mattsson14a} was based on a somewhat speculative modified destruction rate, but had the attractive property of explaining the fact that in most local galaxies $\zeta\sim 0.5$ \citep{Inoue03,Draine07}. Provided there is no dust if $Z=0$, the \citet{Mattsson14a} equilibrium model for an evolved system reduces to $0 = {\zeta} (1-2\zeta)$, which corresponds to $\zeta =  1/2$. A similar balance between growth and destruction has also been found by \citet{Hirashita11}. The model of galactic dust evolution presented above, with the new prescription grain growth, is based on a better motivated change of the grain-growth prescription, which turns out to also yield $\zeta =  1/2$ in the high-$Z$ limit. That is, 
\begin{equation}
\zeta \approx {\epsilon\over \delta + \epsilon} =   {1\over 2}\quad ({\rm if} \,\delta = \epsilon),
\end{equation}
for high $Z$ values, irrespective of $\delta$ and $\epsilon$ (as long as $\delta = \epsilon$) as well as $y_Z$ and $y_{\rm d}$. Fig. \ref{var_yd} shows how $\zeta$ converges to $\zeta = 1/2$ despite different values $y_{\rm d}$, $\epsilon$ and $\delta$, provided that $\delta = \epsilon$. The rate of convergence depends on the actual value of $\delta = \epsilon$, which should be understood as a reflection of the fact that the system is reaching equilibrium faster if growth and destruction is efficient.

\subsection{Comparison with observations}
\label{compobs}
\begin{figure}
  \resizebox{\hsize}{!}{
   \includegraphics{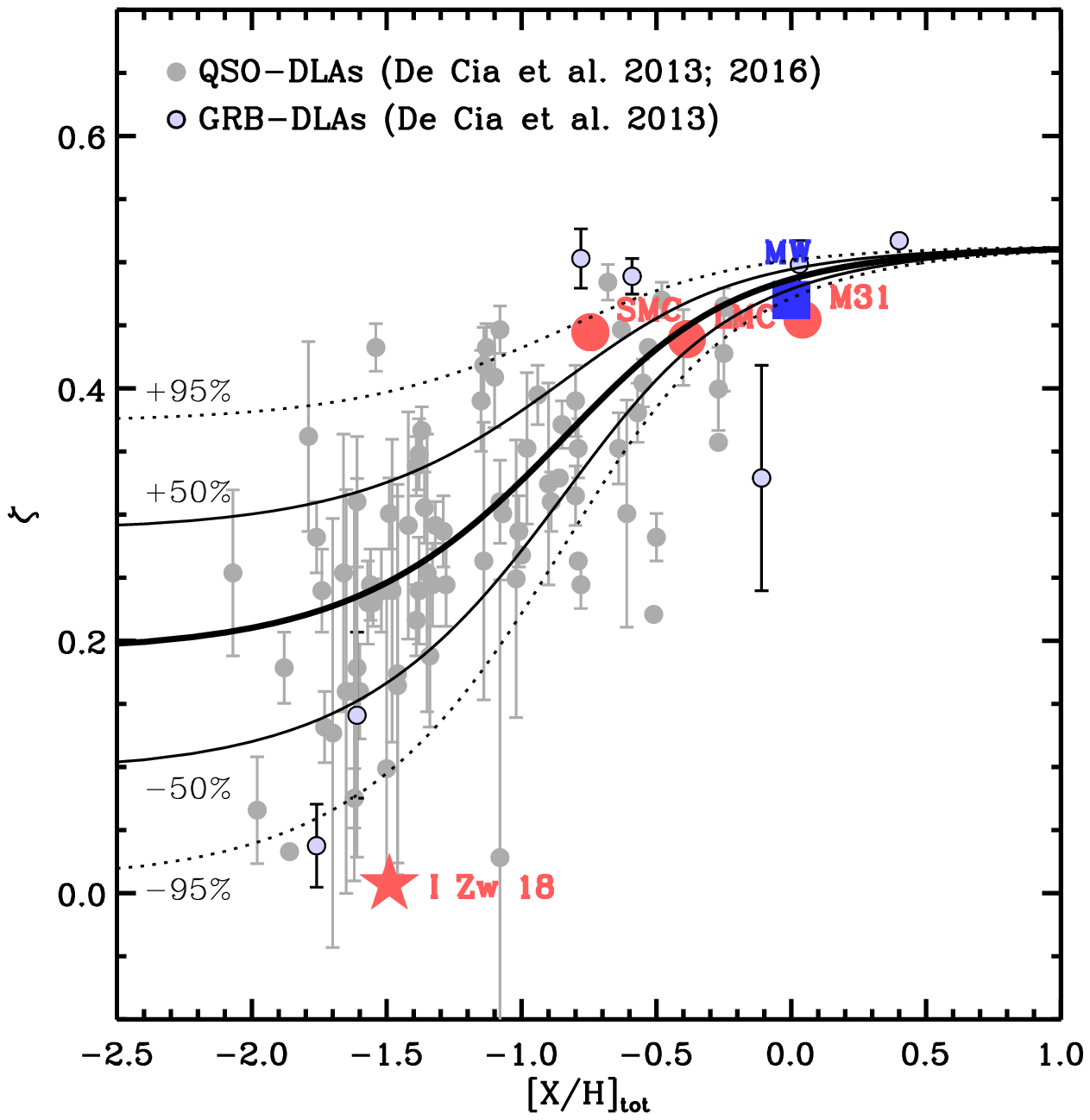}
   }
  \caption{\label{dtm_dep} Comparison with dust-depletion data for a few $\gamma$-ray burst (GRB) damped Lyman-$\alpha$ (DLA) absorbers and a larger sample of DLAs toward quasars (QSOs) taken from \citet{DeCia13} and \citet{DeCia16}, as well as data for the Milky Way, Andromeda, the Magellanic clouds and I Zw 18 \citep{Issa90,Inoue03,Draine07,Herrera-Camus12,Fisher14}. The thick black line shows a model based on a least-squares fit to the data and the thinner black and dotted lines show similar models (best-fit $\epsilon$ and $\delta$) corresponding to $\pm50$\% and $\pm 95$\% deviations from the best-fit value for $y_{\rm d}$. The plot shows the case with $y_Z = 0.01$. The case with $y_Z  = 0.005$ is essentially indistinguishable from this fit.}
  \end{figure}

Qualitatively, the model reproduces the overall trend of $\zeta$ with $Z$ (see Fig. \ref{dtm_dep}) and does so without any {\it ad hoc} modifications of, e.g., $y_{\rm d}$ as in \citet{Mattsson14a}, where an arbitrary dependence of $y_{\rm d}$ on $Z$ was needed because the model curves of $\zeta$ were rising too fast compared to the observational constraints. This issue does not exist with the current model. It also converges towards a single value of $\zeta$ as implied by the observational data around solar and super-solar $Z$. As apposed to the model by \citet{Mattsson14a}, this can now be obtained without modifications of the standard model of interstellar dust destruction. The present model of dust evolution, based on an assumption of fast turbulence-accelerated grain growth in MCs, is therefore simpler, relies on fewer assumptions {\it and} seem to yields a better fit to data derived from observations. 

    \begin{figure}
  \resizebox{\hsize}{!}{
   \includegraphics{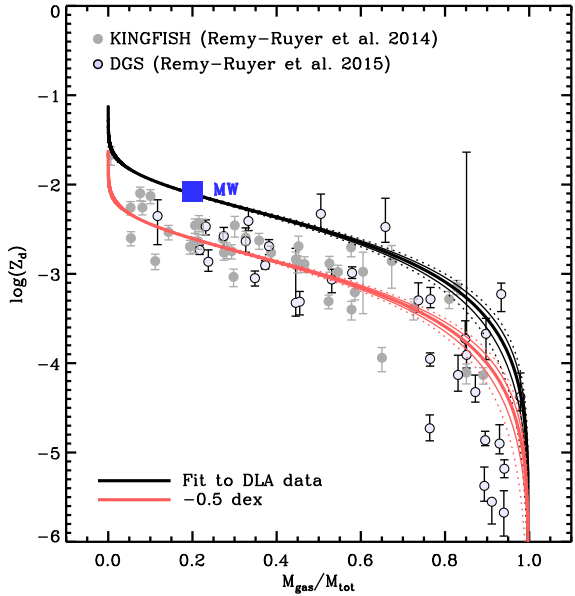}
   }
  \caption{\label{dtg_RR14} Comparison with dust-emission data for local galaxies taken from the KINGFISH and DCG samples of \citet{Remy-Ruyer14} and \citet{Remy-Ruyer15}. The black lines show the model fits to the dust-depletion data  presented in Fig. \ref{dtm_dep} and the red lines show the same model curves with the dust-to-gas ratio scaled down by 0.5 dex. The overall trend implied by the model fits is qualitatively consistent the trend seen in the dust-emission data, albeit with a slight offset in the dust-to-gas ratio. }
  \end{figure}  

To obtain constraints on the free parameters of the model, primarily $\epsilon$ and $\delta$, the solution (equation \ref{solution}) must be calibrated against observational data. Direct least-squares fitting against dust-depletion data for a few $\gamma$-ray burst (GRB) damped Lyman-$\alpha$ (DLA) absorbers and a larger sample of DLAs toward quasars (QSOs) taken from \citet{DeCia13} and \citet{DeCia16}, as well as data for the Milky Way, Andromeda, the Magellanic clouds andI Zw 18 \citep{Issa90,Inoue03,Draine07,Herrera-Camus12,Fisher14}, favours $\epsilon \approx \delta$. Two fits were made: one assuming a fixed ``standard value'' $y_Z = 0.01$ and one with a reduced value $y_Z = 0.005$ (see Table \ref{parfit} for resultant parameter values). 

With $y_Z = 0.01$, the $y_{\rm d}$ is $y_{\rm d}=0.0019$, which means that only about 1/5 of the metals expelled by stars enter the ISM in the form of dust grains (unless galactic outflows are significantly biased towards dust loss). The best-fit model is shown by the thick black line in Fig. (\ref{dtm_dep}) together with models corresponding to $\pm50$\% and $\pm 95$\% variations of $y_{\rm d}$ relative its best-fit value. It is noteworthy that I Zw 18 seem to require a very low $y_{\rm d}$ (roughly 1\% of the the best-fit value), which could be interpreted as evidence for strong outflow effects. It would indeed be consistent with the idea of star-bursting dwarf galaxies having stronger outflows than more massive or quiescent galaxies\footnote{It is, however, noteworthy that effective metal yields seem to increase for galaxies in dense environments \citep[e.g.,][and references therein]{Pilyugin17}. A similar effect could be expected also for dust.}. 

In addition to the GRB/DLA data mentioned above, it is worthwhile comparing with dust- and gas-emission data from local galaxies. Because the total metallicity is difficult to determine accurately from emission spectra of galaxies, it is better to compare with the dust-to-gas ratio $Z_{\rm d}$ in this case. Fig. \ref{dtg_RR14} shows $Z_{\rm d}$ as a function of the gas-mass fraction for objects in the KINGFISH and DCG samples of \citet{Remy-Ruyer14} and \citet{Remy-Ruyer15} over plotted with the model fits obtained with the GRB/DLA data. The trend implied by the model fits is qualitatively consistent the trend seen in the dust-emission data, but there is a slight offset ($\sim0.5$~dex) towards lower $Z_{\rm d}$ in the data. The red lines in Fig. \ref{dtg_RR14} shows the same model fits scaled down by 0.5 dex. It appears as if the dust masses are systematically underestimated in \citet{Remy-Ruyer14} and \citet{Remy-Ruyer15}, because the Milky Way falls on the trend suggested by the GRB/DLA data (the black model lines). However, one should not draw any conclusions from this, as there are many uncertainties involved -- both in the models and in the conversion of observations into physical quantities.

  \begin{table}
  \begin{center}
  \caption{\label{parfit} Resultant fitting parameters from least-squares fitting against dust-depletion data for two different stellar mass yields.}
  \begin{tabular}{l|ll}
  \hline
  \hline
  \rule[-0.2cm]{0mm}{0.6cm}
   $y_Z$ & $0.01$  & $0.005$ \\
  \hline\\[-2mm]
   $y_{\rm d}$ & $1.91\,10^{-3}$ & $9.57\,10^{-4}$ \\
   $\epsilon$   & 4.16  & 2.08 \\
   $\delta$ & 3.95 & 1.97\\[1mm]
  \hline
  \hline
  \end{tabular}
  \end{center}
  \end{table}

The preferred value $\delta \approx 4$ is indicating a destruction rate due to SNe which is consistent with the Milky Way $\tau_{\rm d}$, estimated to be roughly 0.7 Gyr \citep{Jones96}. The effective gas-consumption rate in the solar neighbourhood is about $2\,M_\odot$~pc$^{-2}$~Gyr$^{-1}$, and the gas density is $\sim 8\,M_{\sun}$~pc$^{-2}$ \citep[see, e.g.,][and references therein]{Mattsson10a}, which implies $\delta\approx 5$. The mass of interstellar gas effectively cleared of dust is believed to be in the range  $\langle m_{\rm ISM}\rangle = 500-1000\,M_{\sun}$. Adopting a \citet{Salpeter55} IMF and $\langle m_{\rm ISM}\rangle = 750\,M_{\sun}$ yields $\delta \approx 4$, which suggests that the preferred fitting value is fully consistent with the expected rate of dust destruction. 

There are neither strict, nor independent, constraints on $\epsilon$, which could in principle take any positive value. However, since $\delta$ can be constrained, as argued above, it is very interesting that the best fit ($\epsilon = 4.16$) is so close to the special case $\epsilon = \delta$. But this is no coincidence. As noted above, $\zeta$ for objects of around solar metallicities is $\zeta \approx 0.5$ with a seemingly small statistical scatter (but the number of data points is too small to say anything conclusive). Thus, with data implying $\zeta \approx 0.5$, the fitting algorithm is forced to produce a $\epsilon \approx \delta$ solution. 

As mentioned above, a fit with $y_Z = 0.005$ (the ``standard yield'' lowered by 50\%) was also made. The resultant fit is of the same quality, although the best-fit values for $y_{\rm d}$, $\epsilon$ and $\delta$ are basically also reduced by 50\% compared to the fit with $y_Z = 0.01$. Thus, the prediction that 1/5 of the metals are expelled by stars enter the ISM in the form of dust remains.

\subsection{A note on the correlation between $\epsilon$ and $\delta$ and potential degeneracies}
The preferred model, according to the data mentioned above, suggests a correlation between $\epsilon$ and $\delta$ that is simply $\epsilon \approx \delta$, with the numerical value  $\epsilon \approx \delta \approx 4$ set by the slope of $\zeta$ with respect $Z$ (simple linear regression). As argued above, this value is in agreement with independent estimates of $\delta$, which makes the best-fit model seem very robust. But how reliable is this result, really?

\begin{figure}
  \resizebox{\hsize}{!}{
   \includegraphics{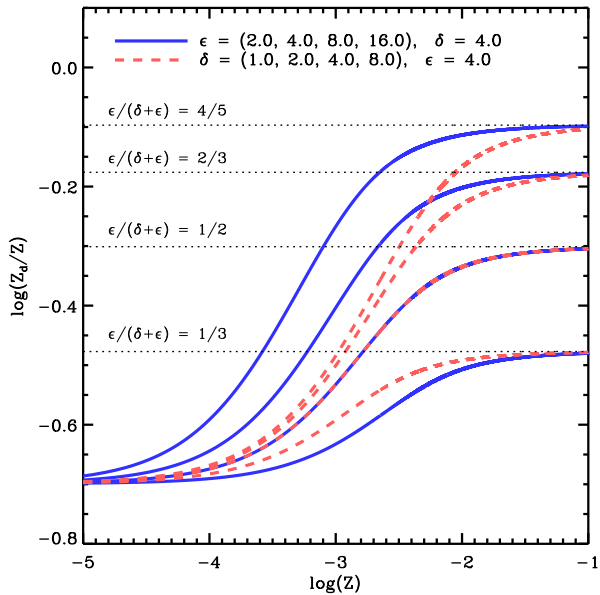}
   }
  \caption{\label{var_epsilon} Dust-to-metals ratio as a function of metallicy. The blue lines show variation of the grain-growth parameter $\epsilon$ with a fixed dust-destruction parameter $\delta$, while the red dashed lines show variation of $\delta$ with a fixed $\epsilon$. Different $\epsilon$ and $\delta$ combinations can result in the same asymptotic dust-to-metals ratio $\zeta_{\rm a} = \epsilon/(\delta+\epsilon)$, which highlights an important parameter degeneracy in the simplistic GDE model considered here. All cases displayed in the figure above assume $y_{\rm d} = 0.002 = 0.2\,y_Z$. }
  \end{figure}

Although the data considered here imply $\epsilon \approx \delta$, one should remember that this is not a universal prediction. It is a result of the fact that there is little scatter in the data at the high-$Z$ end and that $\zeta$ seems to approach $0.5$. If the the asymptotic ratio $\zeta_{\rm a} = \epsilon/(\delta+\epsilon)$ is larger or smaller than the Milky Way value ($\zeta \approx 0.5$) there may be several combinations of $\epsilon$ and $\delta$ which yield the same asymptotic value and display a similar evolutionary track towards that value (see Fig. \ref{var_epsilon} for a few examples). It cannot be ruled out completely that $\delta = 0$ either, but in such a case there must exist an upper limit for the amount of metals accreted onto dust grains in MCs, since $\zeta <1$.

\section{Summary and conclusions}
It has been shown that turbulence can significantly accelerate dust growth by accretion of molecules onto grains small enough to be regarded as coupled to the gas. The growth rate scales with the square of the Mach number, which means that the growth timescale can be reduced by more than an order of magnitude if the Mach number is of the order $10$. This is sufficient to deplete almost all metals in an MC onto dust grains within the lifetime of a typical MC. The limiting timescale is more likely set by the rate of MC formation than the rate of grain growth, which effectively yields a rate of interstellar dust production that roughly follows the star-formation rate. In such a case, dust production in the ISM rapidly reaches the levels needed to explain the dust masses observed at high redshifts ($z = 7-8$) without assuming that SNe are extremely effective dust producers and that the rate of destruction is low.

The growth of dust grains large enough to be regarded as decoupled from the gas flow, is not much accelerated by turbulence (unless coagulation is efficient). On average, there is no difference between the efficiency of growth in a turbulent medium where the gas and the dust are statistically independent,  compared to that of a static uniform medium. Hence, it is concluded that small grains in a turbulent molecular cloud will grow much faster than large grains, which suggests that the GSD will evolve towards formation of a peak around a relatively large grain size. Most of the dust growth is taking place in high-density regions, where the grains rapidly grow to micron size and deplete the growth species. If these large grains then decouple from the gas, the rapid grain growth phase will end.  In a scenario like this, the growth of dust grains is therefore not slow and steady process, but a fast and locally intermittent process.  

Given the results summarised above, can turbulence be the solution to the replenishment problem in models of dust evolution in high-redshift galaxies, i.e., that the regrowth in the ISM is too slow if a ``standard rate'' of dust destruction is assumed \citep[see, e.g.,][]{Mattsson11b,Rowlands14b,Mattsson14a}? The driving of turbulence may be due to shocks originating from SNe, which implies that very high Mach numbers can occur in cold environments where the sound speed is low. An average Mach number $\mathcal{M}_{\rm rms}\sim 10$ can easily be obtained and according to the theory of the present paper, it suggests the grain-growth timescale can easily be reduced by two orders of magnitude. A simple galactic dust-evolution model shows that this is exactly what is needed to maximise dust growth in MCs and compensate for also rather high rates of dust destruction. That is, the elevated SN rates at high redshifts means that SN shocks destroy larger amounts of dust, but the same energy injection by SNe also causes significant turbulence in the cold ISM, which leads to a higher overall rate of dust condensation in the ISM. Previous suggestions of a lower dust-destruction efficiency at early times \citep[e.g.,][]{Gall11a,Gall11b,Mattsson11b,Mattsson14a} may therefore be unnecessary. 
 
With the short dust-growth timescale considered here, the average rate in MCs may be high enough to account for the large dust masses reported in galaxies as early as at redshifts $z\sim 7 - 8$ \citep[see, e.g.,][]{Watson15,Shao19}, even without a large net production of dust from SNe. The latter is important, since there are reasons to believe that much of the dust formed at early stages in SN remnants, will not survive the passage of the reverse shock formed when the blast wave hits the circumstellar and interstellar medium \citep[see, e.g.,][]{Bianchi07,Nozawa07,Kirchschlager19}, although the efficiency of grain destruction depends a lot on the shock velocity and the type of dust \citep{Silvia10}.

As a final remark, it should be noted that results and conclusions of the present paper are obtained based on an idealised model of grain growth in turbulent MCs, involving several simplifying assumptions. Direct numerical simulations of grain growth by accretion of molecules in hypersonic turbulence will be necessary to confirm the theory.

\section*{Acknowledgments}
The author wishes to thank the anonymous reviewer, whose comments, suggestions and criticism led to significant improvement of the original manuscript.
This work is supported by the Swedish Research Council (Vetenskapsrådet), grant no. 2015-04505. 

\bibliographystyle{mnras}
\bibliography{refs_dust}

\label{lastpage}
\end{document}